\documentclass{article}[twocolumn]

\usepackage[english]{babel}

\usepackage[letterpaper,top=2cm,bottom=2cm,left=3cm,right=3cm,marginparwidth=1.75cm]{geometry}

\usepackage{amsmath}
\usepackage{graphicx}
\usepackage[colorlinks=true, allcolors=blue]{hyperref}
\usepackage{microtype}
\usepackage{tabularx} 

\title{Self-supervised Machine Learning Based Approach to Orbit Modelling Applied to Space Traffic Management}
\author{
  Emma Stevenson\textsuperscript{a}, Victor Rodriguez-Fernandez\textsuperscript{a}, \\
  Hodei Urrutxua\textsuperscript{b}, Vincent Morand\textsuperscript{c}, David Camacho\textsuperscript{a} \\[2ex]
  \textsuperscript{a}School of Computer Systems Engineering, Universidad Polit\'ecnica de Madrid, \\
  Calle de Alan Turing, 28038 Madrid, Spain, \\\\
  Email: \{emma.stevenson, victor.rfernandez, david.camacho\}@upm.es \\
  \textsuperscript{b}European Institute for Aviation Training and Accreditation, Universidad Rey Juan Carlos, \\
  Camino del Molino 5, 28942 Fuenlabrada, Spain, \\
  Email: hodei.urrutxua@urjc.es \\\\
  \textsuperscript{c}CNES, 18 avenue Edouard Belin 31400 Toulouse France, \\
  Email: vincent.morand@cnes.fr
}

\date{} 

\begin{document}
\maketitle

\twocolumn

\begin{abstract}


This paper presents a novel methodology for improving the performance of machine learning based space traffic management tasks through the use of a pre-trained orbit model. Taking inspiration from BERT-like self-supervised language models in the field of natural language processing, we introduce ORBERT, and demonstrate the ability of such a model to leverage large quantities of readily available orbit data to learn meaningful representations that can be used to aid in downstream tasks. As a proof of concept of this approach we consider the task of all vs. all conjunction screening, phrased here as a machine learning time series classification task. We show that leveraging unlabelled orbit data leads to improved performance, and that the proposed approach can be particularly beneficial for tasks where the availability of labelled data is limited.

\noindent \textbf{keywords:} self-supervised learning, transfer learning, machine learning, orbit modelling, orbit prediction, conjunction assessment

\end{abstract}


\section{Introduction}
\label{sec: intro}


Ensuring the safety and sustainability of space operations in the New Space era is an ever-increasing challenge for Space Traffic Management (STM) \cite{muelhaupt_space_2019}. In the face of rising space traffic, large constellations, and a growing space debris population, STM activities such as collision avoidance are critical for preserving both current day space assets, and the future usability of the space environment. To address the challenges posed by the scale and complexity of these activities, one emerging approach is the exploitation of recent advancements in the fields of machine learning (ML) \cite{vasile_iac_2017, sanchez_evidencetheory_2020, stevenson_artificial_2021}. 

Advancements in this field are far reaching, with breakthroughs in a variety of different domains benefiting the space sector, from image-based computer vision to text-based Natural Language Processing (NLP). Current applications range from vision-inspired tasks such as space object characterisation \cite{furfaro_stm_pca_2019} and satellite pose estimation \cite{kisantal_satellite_2020}, to the use of NLP techniques for aiding in early space mission design \cite{berquand_space_2020}. However, the success of ML in many of these tasks relies on large, labelled datasets, the availability of which can be a limiting factor in their performance.

Until recently, the achievements of NLP in leveraging vast quantities of unlabelled data have been largely ignored outside of the text realm. In this work, we take inspiration from these techniques, constructing an orbit model that is able to leverage large quantities of readily available orbital data, which can be built upon to perform better STM.

\begin{figure}[h!]
    \centering
    \includegraphics[width=0.9\linewidth]{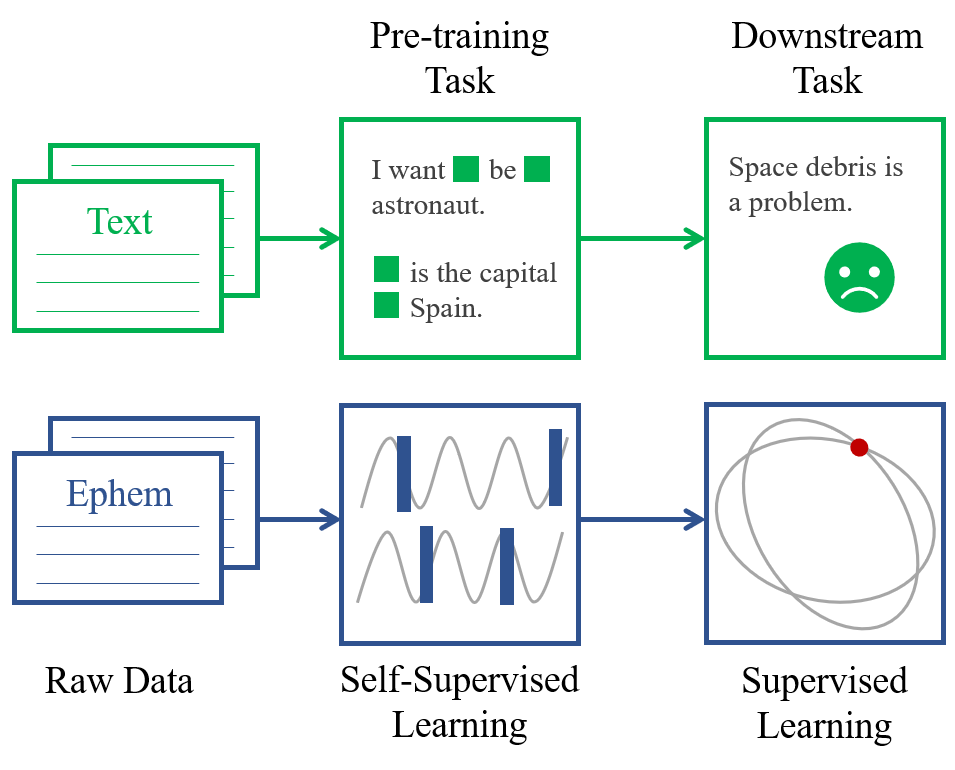}
    \caption{Analogy of the approach proposed in this paper with that used in the domain of NLP. Pre-training a language model to predict missing words or phrases from a large corpus of unlabelled text, for example harvested from wikipedia, enables the model to learn meaningful language representations, which can then be used for downstream tasks such as sentiment analysis, to predict whether a given phrase is positive or negative. Equivalently, pre-training an orbit model based on large quantities of (widely available) unlabelled orbit data, or ephemerides, can be used to learn meaningful orbit representations which can then improve the performance of downstream STM tasks.}\label{fig:nlp_analogy}
\end{figure}

In much the same way that different STM tasks rely on our ability to accurately model orbits, different NLP tasks, such as next word prediction or sentiment analysis, rely on an underlying common understanding of how to model language. The latest breakthroughs in this field can be attributed to the use of self-supervised learning (SSL), whereby high performance underlying language models such as Google’s BERT \cite{devlin_bert_2019} can be pre-trained on extensive datasets by predicting masked words from text, before being fine-tuned to the objectives and data of specific downstream tasks, thus improving both performance and efficiency. Here, we present our proposed approach for applying this concept to the STM domain, and introduce our pre-trained orbit model ORBERT, as well as its application to the downstream task of conjunction screening. This concept is illustrated in Figure \ref{fig:nlp_analogy}, although we stress that such a model could be used for a variety of different downstream applications.

The ORBERT model is trained using a self-supervised approach for time series \cite{zerveas_transformer-based_2020}, where, instead of masking words or phrases, sections of orbit ephemeris are masked, and the model tasked with their reconstruction. This enables it to learn meaningful orbit representations which can be passed to orbit-related tasks such as pairwise conjunction screening, phrased here as a time series classification task. To train these models, we employ a realistic dataset, composed of ephemerides generated by precise numerical propagation of two-line element set (TLE) data, and conjunctions generated using the CNES BAS3E space surveillance simulation framework \cite{morand_bas3e}.

In this paper, we present the capacity of such a self-supervised approach for orbit modelling, as well as the improvement in performance gained by utilising this pre-trained model in the downstream task of conjunction screening. 


The paper is structured as follows. In Section \ref{sec: ML concepts}, we provide backgrounds on the ML concepts relevant to the approach proposed in this paper and define key terminologies for aiding readers unfamiliar with the field. In Section \ref{sec: dataset}, we provide a description of the dataset we employ, before presenting the ORBERT model itself in Section \ref{sec: orbert}. In this section we describe the self-supervised approach, the strengths and weaknesses of the resulting model, and highlight the power of the proposed methodology by giving insights into the learning capacity of the model through an analysis of the learnt representations. Downstream applications of the model, with a focus on the task of orbital conjunction screening, are discussed in Section \ref{sec: downstream}, in which we demonstrate how utilising such a pre-trained model to leverage unlabelled data can be used to improve performance. Finally, in Section \ref{sec: conclusions} we discuss the conclusions of the paper, and outline a variety of promising avenues for future research.

\section{ML Concepts}
\label{sec: ML concepts}

The goal of Machine Learning (ML) is to build a system that can recognise patterns in known input data ($x$), which can then be used to make predictions ($y$) on new data. This field is typically divided into supervised and unsupervised learning: supervised learning refers to the use of a dataset for which both training data and labels exist, whereas unsupervised methods do not have access to labels, but rather seek to uncover the characteristics of different parts of the data set by themselves.

In recent years, many of the breakout successes in ML can be attributed to the subfield of Deep Learning (DL), which uses multi-layer neural networks, often known as Deep Neural Networks (DNNs). Each of these layers, consisting of a set of artificial neurons, takes knowledge extracted from the previous layers and gradually refines it. These layers are laid out following a specific architecture, and are trained by optimisation algorithms to minimise their errors (loss) and improve their accuracy. The capacity of these DNN architectures for solving complex tasks such as image classification using large labelled datasets, namely ImageNet, has made the field focus more intensively on supervised problems with labelled data.

However, creating a dataset with a sufficient number of annotated examples is a challenging task, and labeling data is arduous, expensive, and sometimes infeasible. In this sense, self-supervised learning is an exciting research direction that aims to alleviate this problem by learning representations using labels that are embedded in the data itself, without explicit and potentially even manual supervision. Many of these methods are developed in specific communities such as natural language processing, computer vision or graph learning \cite{liu2021self}. One of the major benefits of self-supervisory learning is the ability to scale to large amounts of unlabelled data in a lifelong learning manner.

Self-supervised learning is not usually used for the model that is trained directly, but instead is used as a ``pretext task" that pre-trains the model before being updated (or ``fine-tuned") on a final task, commonly known as the downstream task. This process is known as transfer learning, and constituted a major milestone for deep learning by enabling researchers with few resources to train effective models more quickly and with less data. 

Although there are many techniques for pre-training models in a self-supervised way, in this work we draw inspiration from the Masked Language Model (MLM) methods used in Natural Language Processing (NLP). Here, the input sentences from a text dataset are randomly masked, and the aim of the model is to recover the masked word (See Figure \ref{fig:nlp_analogy}). This technique gave rise to BERT \cite{devlin_bert_2019}, a breakthrough language model that was pre-trained using an enormous dataset, and that could be applied to a variety of different downstream tasks such as sentiment analysis (See Figure \ref{fig:nlp_analogy}), achieving state-of-the-art performance without having been trained for that task directly. Although BERT is based on a popular neural architecture known as the Transformer, the technique with which it has been pre-trained can be applied using any neural architecture. 

\begin{figure}[h!]
    \centering
    \includegraphics[width=0.9\linewidth]{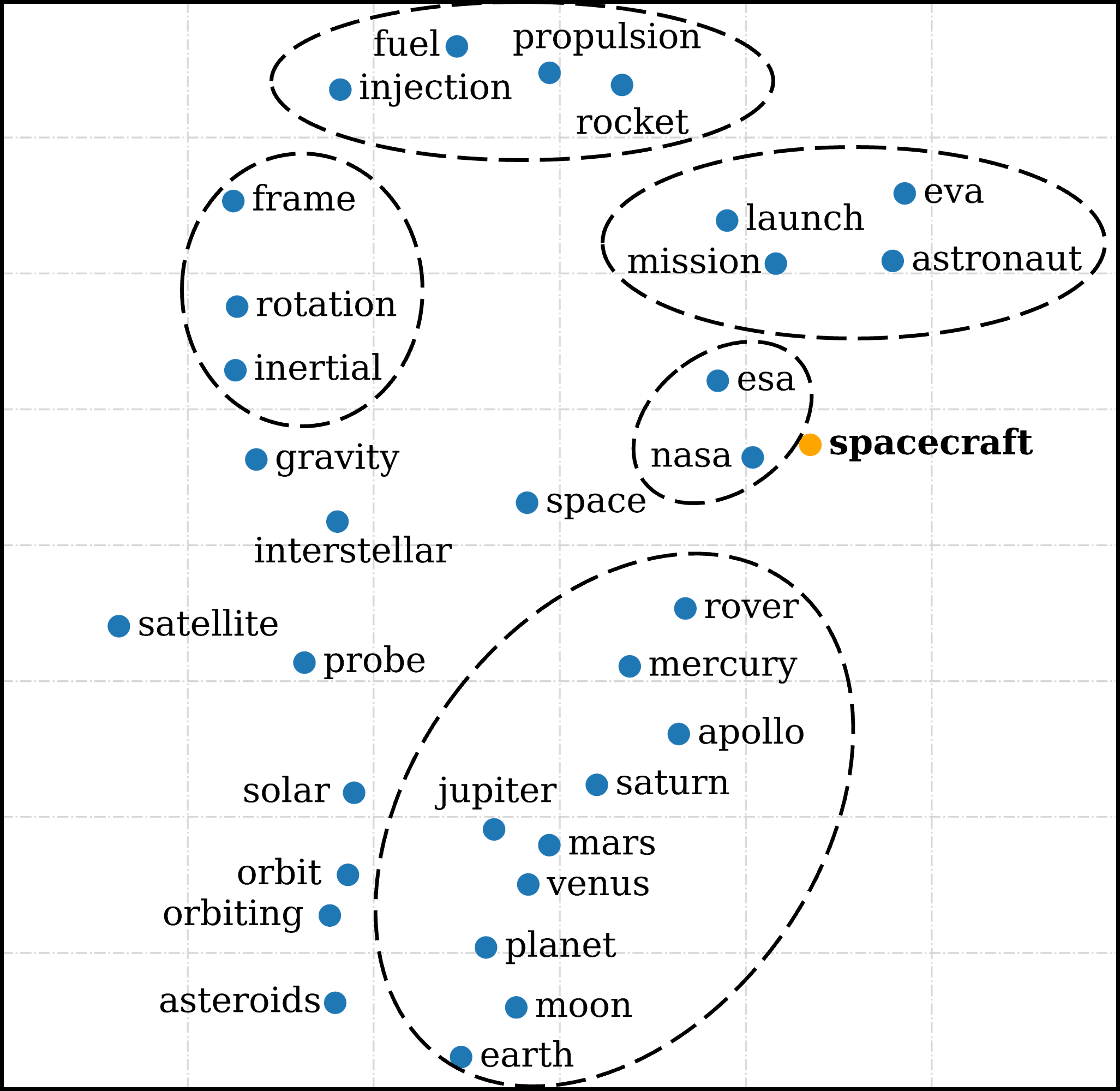}
    \caption{Illustrative example of the concept of clustering in extracted model representations (embedding space) from the world of NLP. Word embeddings extracted from the Word2vec model \cite{smilkov_embedding_2016}, and selected here as certain nearest neighbours of the word \textit{spacecraft}. Clear examples of clusters of semantically similar words are highlighted: for example, the planets which a spacecraft might visit are close in embedding space, and separated from those related to the launch, or rocket.}\label{fig:word2vec_clustering}
\end{figure}

One interesting aspect of BERT and other self-supervised models lies in analysing the content of the neurons, also known as activations or embeddings, for different samples of the dataset in the later layers of the neural network, which often reveal how the model ``understands" the data. One common approach to do this consists of projecting the activations of each sample into a 2D or 3D space through a dimensionality reduction technique, and then running a clustering algorithm to group together similar data points. A popular example of this can be found, again, in the field of NLP, where projecting the activations of word-based pre-trained models shows how words that are semantically similar appear close in the projected space, and vice versa (See Figure \ref{fig:word2vec_clustering}). This level of knowledge and abstraction emerges from the model simply by training it with enough data in a simple self-supervised pretext task such as predicting masked words.


\section{Dataset} 
\label{sec: dataset}

In this section, we describe the generation of the datasets we employ for training the models in this work. The first of these, which we use to pre-train the orbit model in a self-supervised fashion, is an unlabelled dataset consisting solely of orbit ephemerides. From this, we then generate a second, labelled dataset to be used in the training of the downstream task, chosen here as that of conjunction screening. As we aim to show that the power of this approach lies in its ability to leverage unlabelled data for tasks where the labelling process may be expensive, for the second dataset we consider the computationally intensive \textit{all vs. all} conjunction scenario. In this scenario, all possible pairs of catalogued objects, both active and debris, are screened for conjunctions.

We generate these datasets using BAS3E (\textbf{B}anc d'\textbf{A}nalyse et de \textbf{S}imulation d'un \textbf{S}ysteme de \textbf{S}urvei\-llance de l'\textbf{E}space – Simulation and Analysis Bench for Space Surveillance System), a CNES-owned tool developed in JAVA whose objective is to model and estimate the performances of space surveillance systems, both on ground and on board. BAS3E capabilities are wide and include space object population orbit propagation, sensor modelling and scheduling, correlation, orbit determination and lately simulation of services such as the collision avoidance services, all offered through a modular approach \cite{morand_bas3e}.

We first import an initial population of TLEs, taken as that of the 1\textsuperscript{st} of June 2020, retrieved from space track. Only Low Earth Orbit (LEO) objects with an altitude under 2000~km are kept, resulting in an orbit dataset comprising 18415 objects. We then propagate this population over a 7-day period using the force model detailed in Table \ref{tab:forcemodel}.

\begin{table}[h!]
  \begin{center}
    \caption{Force model used to propagate the TLE population.}\vspace{1em}
    \renewcommand{\arraystretch}{1.2}
    \begin{tabular}[h]{p{3cm}p{4cm}} 
      \hline
      \textbf{Perturbation} & \textbf{Model for population propagation} \\
      \hline
      Earth potential  & WGS84 Earth model with 12x12 development \\
      Atmospheric Drag  &  MSIS00 atmospheric model with constant solar activity ($F10.7=140$ sfu, $Ap=9$) \\
      Solar radiation pressure & Cannonball model (Earth eclipses considered) \\
      3\textsuperscript{rd} body perturbation & Sun and Moon considered \\
      \hline \\
      \end{tabular}
    \label{tab:forcemodel}
  \end{center}
\end{table}

It should be noted that for the propagation, the area to mass ratio has been set to a constant value of 0.01~m\textsuperscript{2}/kg for all objects, ignoring the Bstar value from the TLE.

Using the orbit definitions given in \cite{esa_envreport_2021}, the resulting orbit dataset consists of 14188 Low Earth Orbit (LEO) objects, 477 Medium Earth Orbit (MEO) objects, 1171 LEO-MEO Crossing Orbit (LMO) objects, 545 Geostationary Transfer Orbit (GTO) objects, 806 Geostationary Orbit (GEO) objects, 121 Highly Eccentric Earth Orbit (HEO) objects and 1107 undefined orbits. 


The all vs. all dataset is then generated by checking every possible pair of the orbit dataset for conjunctions over the propagation timespan (chosen to be representative of typical conjunction screening periods for LEO). Considering 18415 objects, this evaluation is performed for 170 million object pairs. Using a distance-based threshold of 20~km as the conjunction criterion, and considering multiple encounters (repetitive conjunctions between same objects), we detect 1.5 million conjunctions over this one-week period. For each conjunction, the objects involved, as well as the time of closest approach (TCA) and distance of closest approach, are included in the dataset (for more details on the generation of the all vs. all dataset, we refer the reader to \cite{stevenson_artificial_2021}).

This computation was performed on the CNES High Performance Computing (HPC) service, taking a total of approximately 4 days of wall time. It should however be noted that this includes the waiting time to access resources on the shared cluster of machines. Although the BAS3E algorithms were not optimised to perform such an analysis, this very high compute time for generating a labelled dataset for a single snapshot of the space object population demonstrates the interest of developing methodologies for leveraging widely available unlabelled orbit data. 

\section{Orbit Model}
\label{sec: orbert}

Inspired by the breakthrough success of BERT-like masked language models in NLP, we follow the naming convention taken by many of its derivatives (from ALBERT to RoBERTa), and in this section introduce our pre-trained orbit model, ORBERT. Unlike its text-based cousins, we train ORBERT using ephemeris data by translating the masking concepts used in NLP to the time series domain. For this, we apply an analogous approach where, instead of masking words or phrases, sections of the time series are masked, and the model tasked with their reconstruction \cite{zerveas_transformer-based_2020}. In this way, the model learns using a self-supervised approach (as discussed in Section \ref{sec: ML concepts}), using labels obtained from the data itself. This approach enables to pretrain a model that leverages potentially large quantities of unlabelled training data, extracting meaningful vector representations of the time series which can subsequently be used for a variety of downstream time series tasks such as regression, classification and forecasting.

In Section \ref{sec: orbert: method}, we detail this masking and self-supervised training approach, for which we employ the \emph{tsai} library \cite{tsai}, based on the deep learning framework \emph{fastai}. In Section \ref{sec: orbert: results}, we present some strengths and weaknesses of the resulting ORBERT model, and finally, in Section \ref{sec: orbert: insights}, we perform a detailed cluster analysis based on the learnt vector representations to give insights into the learning capacity of the model, and its potential for aiding in downstream orbit-related tasks. 

\subsection[Self-Supervised Training Approach]{Self-Supervised Training \\ Approach} 
\label{sec: orbert: method}


Employing this concept of self-supervised learning for time series \cite{zerveas_transformer-based_2020}, we train the ORBERT model under the following procedure. We treat the orbit ephemerides, whose generation is discussed in Section \ref{sec: dataset}, as 6-channel multivariate time series of sequence length 2016 (a 7-day period with step size of 5 minutes). For this work, we consider a Cartesian data representation, with the state of a given object for each time step $t$, given in the EME2000 reference frame, described by $\vec{X}(t) = [X, Y, Z, V_{X}, V_{Y}, V_{Z}]^{T}$.

For pre-processing, we first divide the orbit dataset into training and validation subsets using a random splitting strategy to ensure that the underlying distribution of both subsets is the same. This validation set is used to independently evaluate the performance of the model during training, essential to ensure that the model is not overfitting, and will generalise well to new data. We use a nominal 80\% to 20\% splitting strategy, assigning 14732 and 3683 objects to the training and validation subsets respectively. We then standardise the data, transforming each variable to have zero mean and unit variance to account for differences in units and scales, and to improve the numerical stability of the model training.

\begin{figure}[h!]
    \centering
    \includegraphics[width=0.99\linewidth]{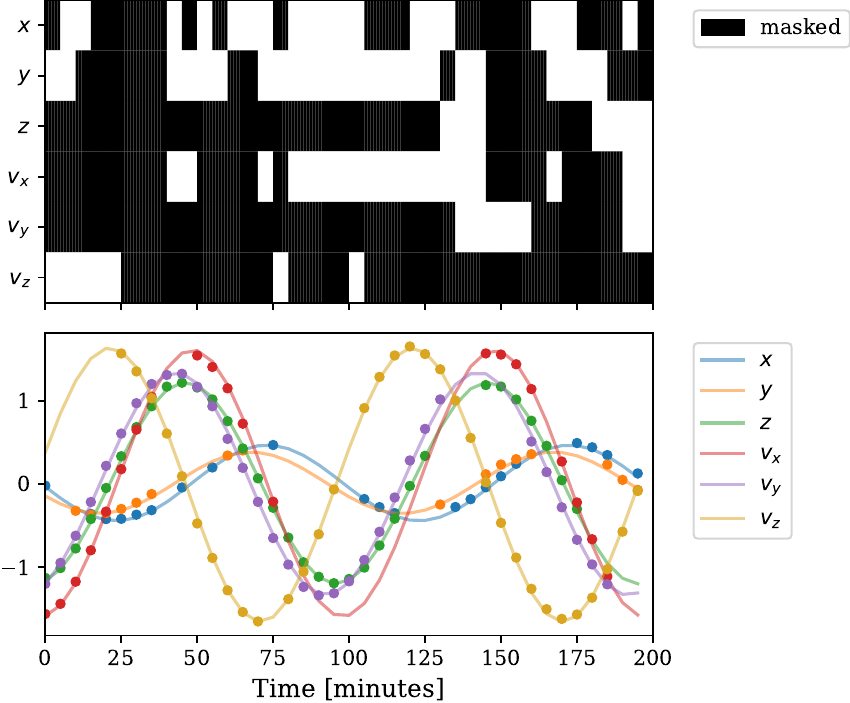}
    \caption{Example of self-supervised masking strategy: mask (top); predicted values over mask denoted by point markers on normalised orbit data (bottom). }\label{fig:mask_eg}
\end{figure}

Next, we apply the mask to the data. This is applied stochastically over the time series, controlled by the probability of masking, $r$, and the mean length of the masked segments, $lm$. The latter variable is required in order to force the model beyond trivial prediction methods such as padding, replication or linear interpolation, which might offer a sufficiently good approximation for very short masked sequences. Furthermore, we elect not to use the same mask synchronised over all variables for a given object, but instead generate a new mask for each in order to encourage the model to learn both relationships between different values along individual sequences, as well as inter-dependencies between variables in order to improve the modelling. A binary sequence is generated based on these parameters (given in Table \ref{tab: mask settings}) for each channel, and the model tasked with predicting the values of the time series over which the mask is set to 0. This concept is illustrated in Figure \ref{fig:mask_eg}.

\begin{table}[h!]
  \begin{center}
    \caption{Hyperparameter settings for the self-supervised masking procedure.}\vspace{1em}
    \renewcommand{\arraystretch}{1.2}
    \begin{tabular}[h]{ll}
      \hline
      \textbf{Parameter} & \textbf{Value} \\
      \hline
      Probability of masking, $r$ & 0.5 \\
      Average mask length, $lm$ & 5 \\
      Synchronised masking & False \\
      \hline \\
      \end{tabular}
    \label{tab: mask settings}
  \end{center}
\end{table}

To train the model to make these predictions, we employ the InceptionTime architecture \cite{fawaz_inceptiontime_2020}. This is, to the best of our knowledge, the best deep neural architecture for time series tasks among the family of 1-dimensional Convolutional Neural Networks (CNNs). We chose this family of networks as they are most suitable for capturing short-term patterns in the data, such as orbit oscillations, since they compute features using sliding convolutional filters. In this work, we use a network made up of six sequential inception modules which maintain residual connections, a common strategy to build deeper CNNs. For the output layer, we reconstruct the shape of the input data by using a convolutional layer  with a number of filters equal to the number of channels in the input data, and a filter size of 1.



We train the model using an adaptation of the nominal loss function for regression problems in ML, the Mean Squared Error (MSE), such that it only considers the predictions over masked values. For a given object with $i$ channels and $t$ time steps, this loss is defined as,

\begin{equation}
    \mathcal{L} = \mathop{\sum \sum}_{(t, i) \in M} (\hat{x}(t, i) - x(t, i))^{2},
    \label{eq: loss_func}
\end{equation}

\noindent where $x$ are the true values of the time series and $\hat{x}$ those predicted by the model over the set $M \equiv \{(t,i) : m_{t,i} = 0\}$, where $m_{t,i}$ are the elements of the mask $M$.

For reproducibility, we detail the hyperparameters used to obtain successful convergence of the model in Table \ref{tab: model settings}.

\begin{table}[h!]
  \begin{center}
    \caption{Hyperparameter settings for the deep learning model architecture and training procedure.}\vspace{1em}
    \renewcommand{\arraystretch}{1.2}
    \begin{tabular}[h]{ll}
      \hline
      \textbf{Parameter} & \textbf{Value} \\
      \hline
      Architecture & InceptionTime \\
      Number of filters & 32  \\
      Dropout & 0.1 \\ 
      Loss Function & Masked MSE \\
      Optimiser & Adam \\
      Learning Rate & 0.003 \\
      Epochs & 100 \\
      Batch Size & 64 \\
      \hline \\
      \end{tabular}
    \label{tab: model settings}
  \end{center}
\end{table}

\subsection{Model Performance} 
\label{sec: orbert: results}

When training a pretext model, it is not the final performance of this intermediate task that is important, but rather we are interested in the learnt representations and whether this can carry meaningful, valuable information downstream. For example in the case of ORBERT, which is effectively trained on an imputation task, we do not intend to directly apply this model to predict missing values in orbit ephemerides, but rather to use the model as a foundation for improved ML-based STM applications. 

\begin{figure}[h!]
    \centering
    \includegraphics[width=0.95\linewidth]{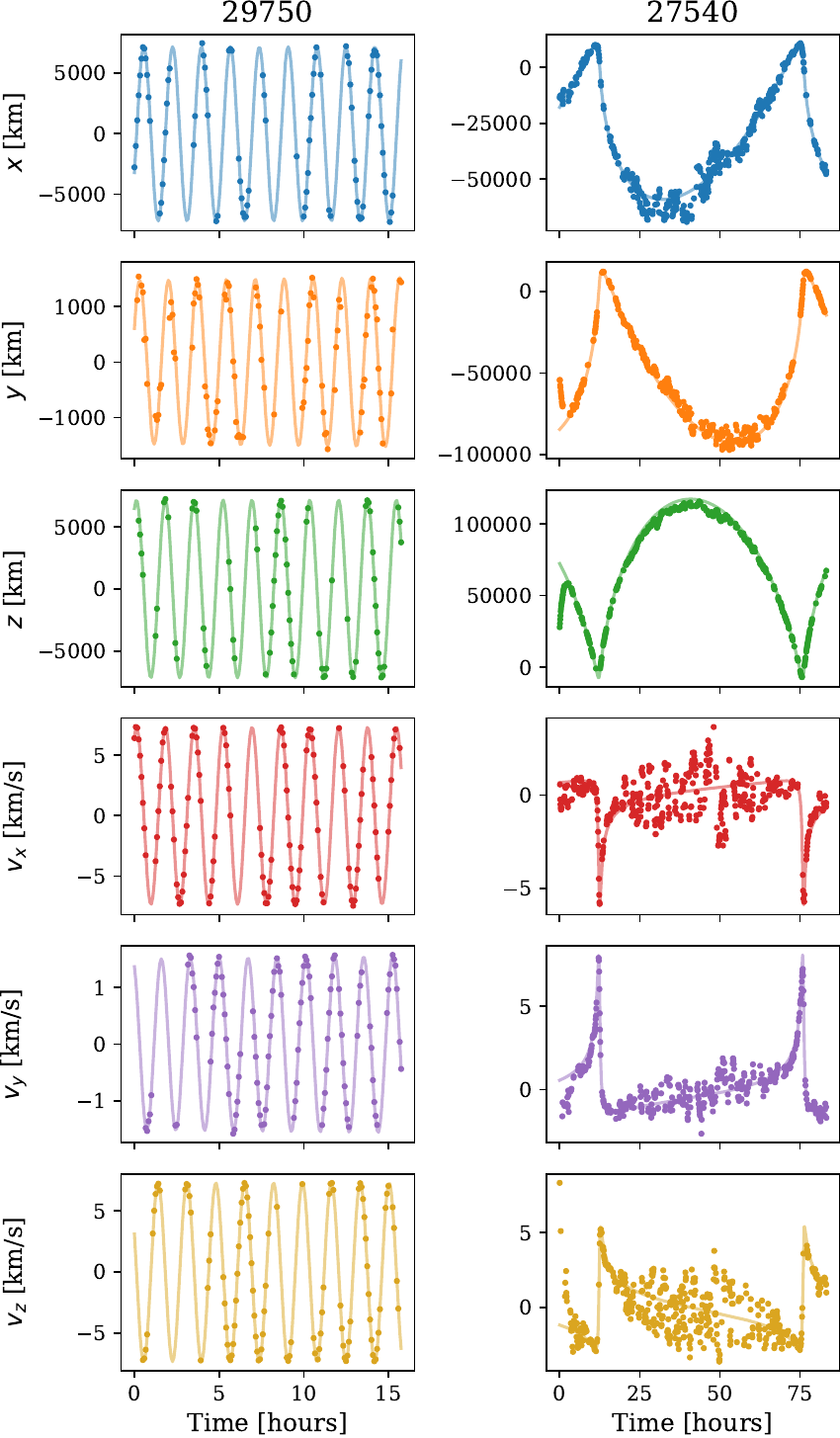}
    \caption{Examples of model top losses. On the left, the object for which the model performed best (lowest loss), NORAD ID 29750, Fengyun 1C debris (LEO). On the right, the object for which the model performed worst (highest loss), NORAD ID 27540, INTEGRAL (HEO). Predicted values of the model over mask are denoted by point markers, and time series have been truncated to highlight important trends.}\label{fig:top_losses}
\end{figure}

It is therefore important to see the strengths and weaknesses of the model, which may translate downstream. In Figure \ref{fig:top_losses}, we show the best and worst performing cases of the model on the validation set. We found that the model performed particularly well for LEO-altitude, low eccentricity objects such as that shown on the left, and systematically badly for high eccentricity objects, as shown on the right. Notably, it can be seen at low timestamps for the $X$ and $Y$ components of the HEO object, and intermediate timestamps for the velocity components, that the model is attempting to fit a higher frequency periodic motion. 

Foremost, this is a natural consequence of the bias present in the original dataset, which is comprised predominantly of LEO objects, and very few HEO-type objects (as detailed in Section \ref{sec: dataset}). Such an effect could therefore be remedied by more careful construction of the training set if such orbital regimes are important for the downstream task. However, we also found that the model appeared to perform better when using a Cartesian data representation, as shown here, as opposed to a Keplerian representation $\vec{X}(t) = [a, e, i, \omega, \Omega, \nu]^{T}$ (which also had the added complexity of requiring a circular loss function for angle variables). This suggests that the modelling approach is better able to capture a smooth varying sinusoidal motion for which the period is the same over all channels, and will be investigated further in future work. 


\subsection{Model Insights: Extracted Representations}
\label{sec: orbert: insights}



To give insights into whether a pretext model has learnt meaningful representations of the data that can be used to aid with downstream tasks, we can examine the geometry of the embedding space (as discussed in Section \ref{sec: ML concepts}). 

To explain how this information is extracted, we can consider a simple neural network, which consists of an input layer, a series of hidden layers, and an output layer, where the knowledge extracted by the network is compressed into the final predictions. The bulk of the network therefore acts as a feature extractor, and thus in order to examine the hidden knowledge distilled by the network, we want to examine the final hidden layer, before this knowledge is compressed. Each of the neurons in this layer will be activated by different aspects, or features, of the data. As such, when a given orbit is passed through this section of the network, the neurons associated with the strongest or most relevant features for that orbit will be more activated. To interpret the representations that a model has learnt, we can therefore examine the activations of these neurons for all orbits.

However, these high-dimensional feature vectors are difficult to visualise, and thus we can use dimensionality reduction techniques to translate these large sparse vectors into a lower-dimensional space in such a way that the compressed representation retains meaningful properties of the original data. In the text analogy, this process preserves so-called semantically similar relationships, for example, as can be seen in Figure \ref{fig:word2vec_clustering}, keeping \textit{Mars} close to \textit{Venus} but separating it from \textit{rocket}. Similarly, we might expect orbits with similar properties to be clustered together. By analysing these geometric relationships and patterns, we can therefore gain important model insights, and assign a level of interpretability, explainability and confidence to the model. \cite{smilkov_embedding_2016}

\begin{figure}[h!]
    \centering
    \includegraphics[width=0.8\linewidth]{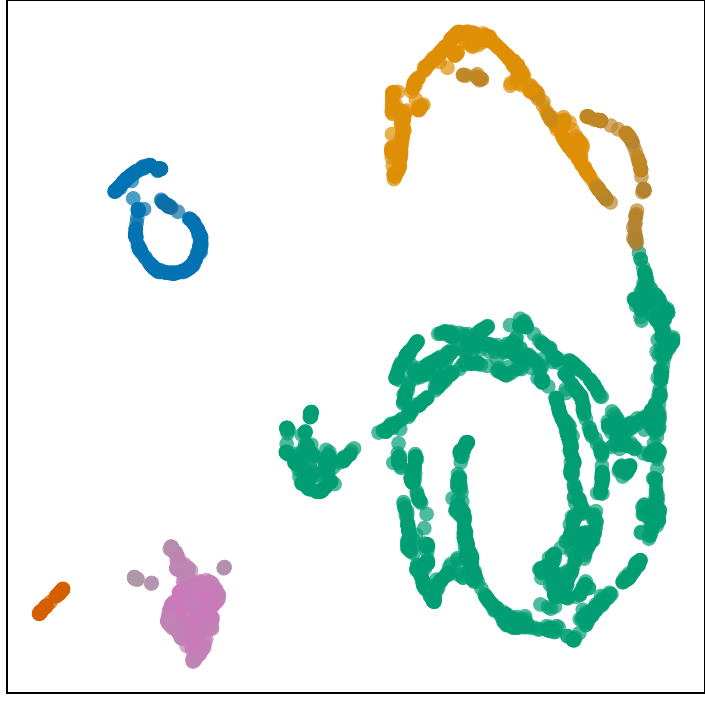}
    \caption{ORBERT two-dimensional orbit embeddings colour-labelled as belonging to identified clusters.}\label{fig:clusters}
\end{figure}

Using the approach described above, we extracted the activations of the final layer of the InceptionTime architecture used for training ORBERT for the orbits in our validation set. For each object, we therefore obtain activations for each neuron at each timestep over the sequence length. To reduce the dimensionality of the vector space, we elected to use mean pooling, averaging the activations over the sequence length, before employing three commonly used dimensionality reduction techniques: Principal Component Analysis (PCA), T-distributed stochastic neighbour embedding (t-SNE) and Uniform Manifold Approximation and Projection (UMAP). Of the three, we found UMAP exhibited the best clusters, and therefore chose this representation for applying the clustering algorithm HDBSCAN. The resulting two-dimensional embeddings and clusters are shown in Figure \ref{fig:clusters}.


\begin{figure*}[htb]
    \centering
    \includegraphics[width=0.8\linewidth]{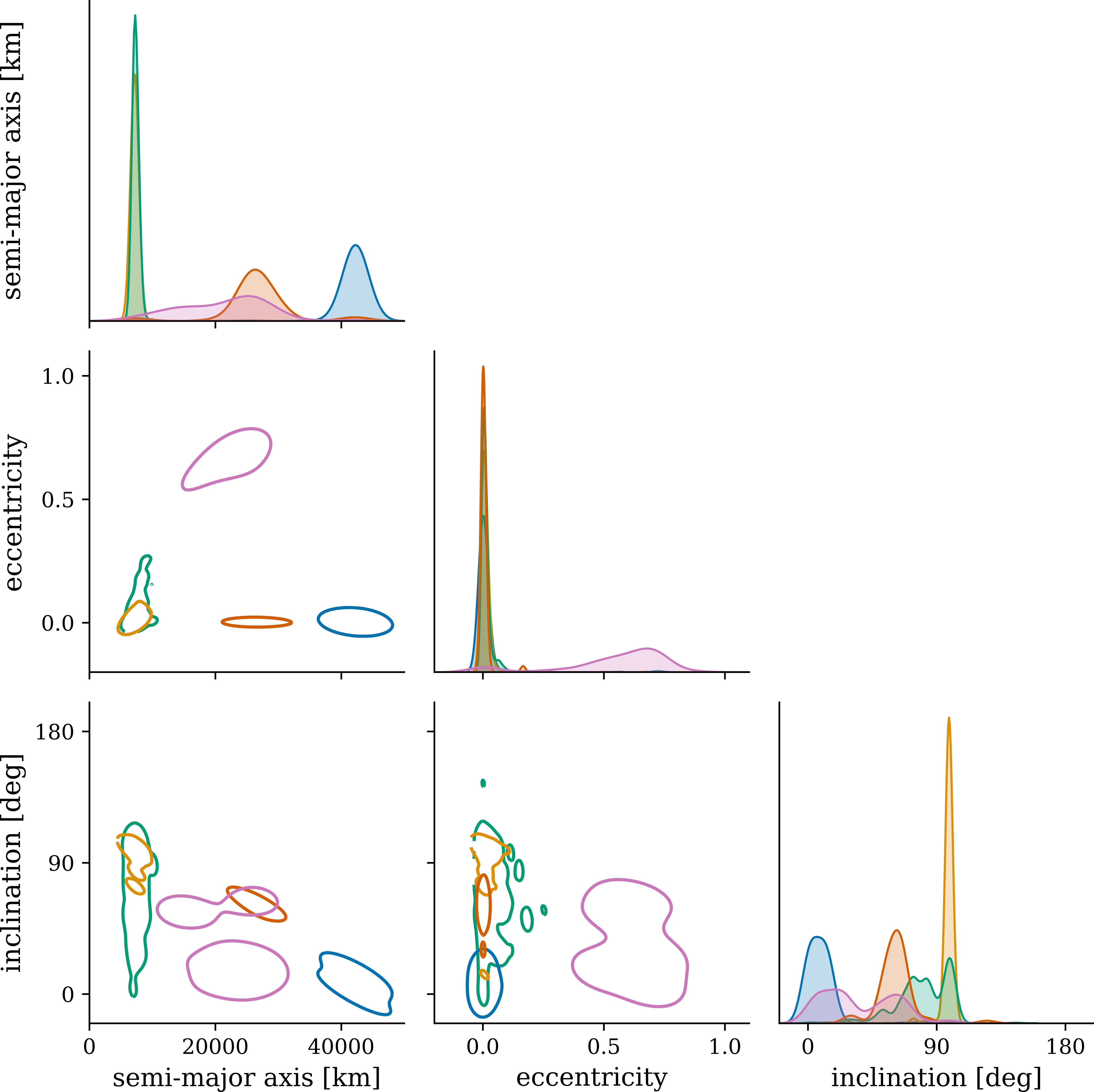}
    \caption{Distributions of orbit properties over the clusters of the ORBERT embedding space (same colour scheme as Figure \ref{fig:clusters}). Diagonal plots are marginalised univariate, and off-diagonal bivariate, distributions plotted using kernel density estimation. Clusters can be seen to represent distinct orbital regimes.}\label{fig:cluster_insights}
\end{figure*}

To investigate whether these clusters have physical meaning, and ensure that the model is behaving as expected and will benefit downstream tasks, we plot the distribution of orbital elements over the objects in each of the identified clusters. This is shown in Figure \ref{fig:cluster_insights} for the most insightful elements, with the marginalised univariate distribution for each element displayed on the diagonal, and bivariate distributions for inter-relations between the elements displayed on the off-diagonal. 

From Figure \ref{fig:cluster_insights}, we can see that the green and orange clusters both belong to the Low Earth Orbit (LEO) altitude regime, but are differentiated by their inclinations, with the orange cluster representing objects in Sun-Synchronous Orbits (SSO) with $98^{\circ}$ inclination. Although these were the most prevalent orbits in the original dataset, we see three further, distinct clusters. The red cluster, with low eccentricity and inclinations of $\sim 55^{\circ} - 65^{\circ}$, corresponds to the Medium Earth Orbit (MEO) region used by navigation satellites, and the blue, with low eccentricity and inclination, to Geostationary Orbit (GEO) type orbits. Finally, the pink cluster encapsulates the behaviour of high eccentricity orbit types, with the higher peak of its bimodal distribution in inclination at $\sim 64.3^{\circ}$ consistent with Molniya type orbits.

We confirm these conclusions by investigating the most representative points for each cluster as returned by HDBSCAN. Unlike commonly used centroid-based clustering algorithms such as k-means, HDBSCAN supports arbitrary shapes for clusters and thus no single point can be returned as ``most representative''. Using the orbit definitions given in \cite{esa_envreport_2021} for the ``exemplar'' points of each cluster, we found the green and orange clusters to be comprised of LEO objects, red MEO, blue GEO and pink LEO-MEO Crossing Orbits (LMO) and Geostationary Transfer Orbits (GTO). Highly Eccentric Earth Orbit (HEO) objects were not found in these sets of most representative points, further explaining the models poor performance in these cases, as seen in Section \ref{sec: orbert: results}.


We can therefore conclude that the model is behaving as expected, with similar orbits clustered together, enabling us to establish trust in the algorithm, as well as giving insights into what it has learnt. In fact, we see that despite learning from a Cartesian representation of the orbit data for an imputation task, the orbit model is able to successfully classify the orbits in Keplerian space. This concept illustrates the power of pre-training, where hidden knowledge learnt from one (pretext) task can passed on to improve the performance of different (downstream) tasks. In the case of ORBERT, this understanding of how space objects move under different perturbation regimes, can be used to improve the performance of a variety of different STM tasks.




\section[Downstream Case Study: All vs. All Conjunction Screening]{Downstream Case Study:\\ All vs. All Conjunction\\ Screening}
\label{sec: downstream}


With an understanding of the orbital motion of space objects at the heart of many STM tasks, the power of such an orbit-based pretext model lies in its ability to aid in a variety of practical ML-based STM applications. Without any modification, we could apply the ORBERT model to missing value imputation, or adapt it to the task of orbit propagation by changing the masking procedure to cover the end sequence of all variables simultaneously. Alternatively, we can utilise the trained weights for downstream time series tasks such as classification or regression, exploiting the extracted orbit knowledge to provide a boost in performance, especially when considering small labelled datasets. In this section, we consider this latter approach for supporting the downstream task of conjunction screening, phrased here as a time series classification task.

At present, space operators typically only screen for possible conjunctions of their own set of active satellites with the full debris field, the \textit{one vs. all} problem. In this approach, an operator can assess the risk of a direct collision, and make a decision as to whether a collision avoidance manoeuvre is required to protect their asset. As such, collision warnings are only raised when one party is a large, active satellite with manoeuvring capability. This leaves millions of unscreened possible debris-debris collisions, whose resulting fragment cloud could cause irreparable damage or even total loss of life to operational satellites. To have a proper awareness and understanding of the space environment as a whole, it is therefore essential to consider collisions between all possible sets of catalogued objects. However, this problem of \textit{all vs. all} screening is not usually addressed as it is extremely computationally expensive, and threatens to become ever-more so in the face of increasing space traffic and observational capabilities in the New Space era. This motivates the need to look for new approaches such as those provided by machine learning  \cite{stevenson_artificial_2021}.

We therefore consider the following case study. An operator routinely performs conjunction assessment for their own fleet in a one vs. all approach, and thus possesses an annotated conjunction/no-conjunction dataset for these pairs over a given screening period. The operator would like to additionally perform routine all vs. all conjunction screening for better space situational awareness, using an ML-based approach for efficiency, but the labelling of an all vs. all dataset for training this model is a prohibitively expensive process. The question is thus: can we leverage unlabelled orbit data to generalise from this limited one vs. all dataset to predict conjunctions in an all vs. all scenario?

\subsection{Dataset}
\label{sec: downstream: dataset}

\begin{figure}[h!]
    \centering
    \includegraphics[width=0.99\linewidth]{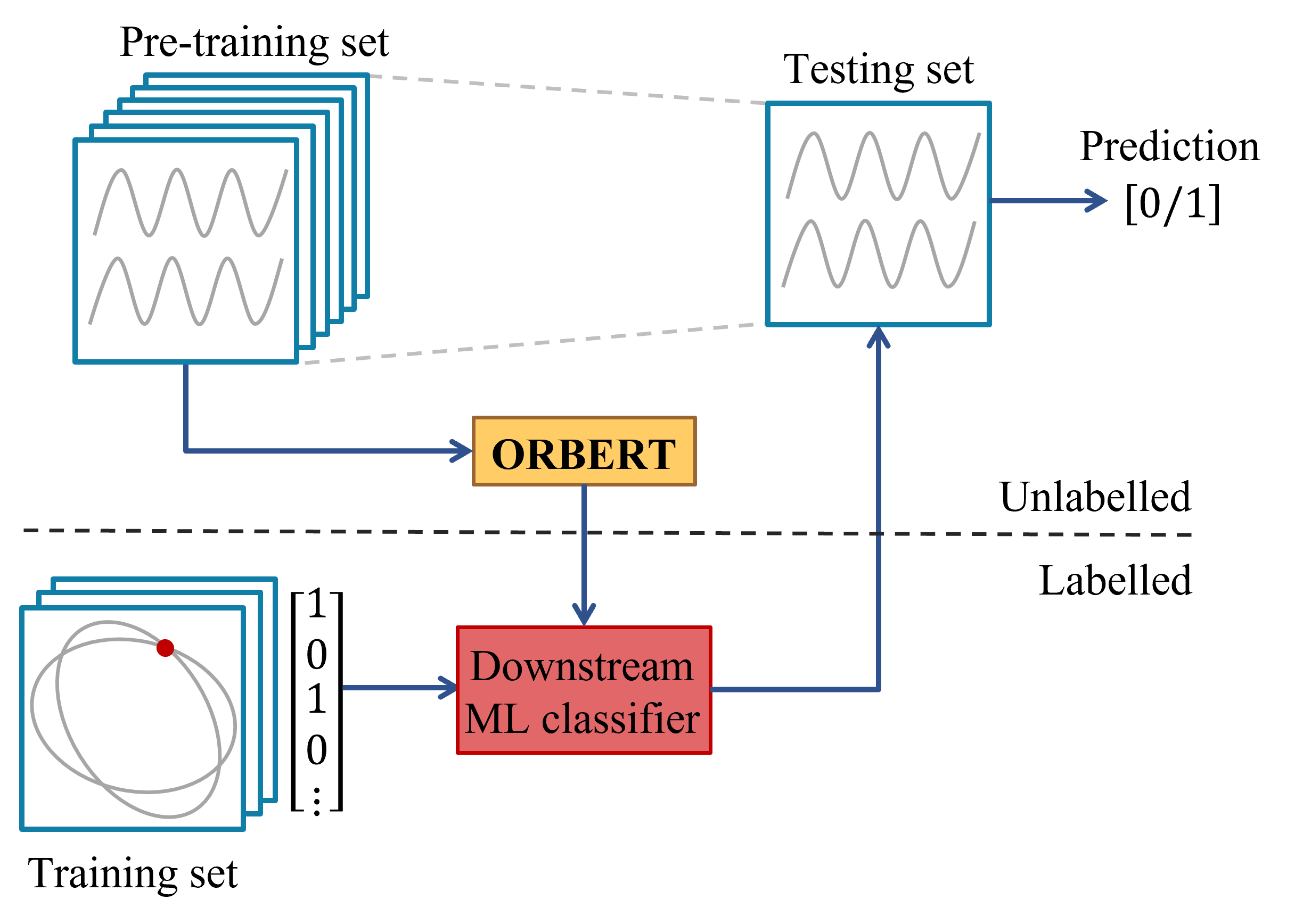}
    \caption{Illustration of the three datasets involved in the downstream experiment.}\label{fig:flowchart}
\end{figure}

To evaluate any performance boost obtained by employing a pretext model, we require three datasets: pre-training (unlabelled), training (labelled) and testing (see Figure \ref{fig:flowchart}). The training data represents the labelled data the operator has for the task, the pre-trained data is all the available unlabelled orbit data, and the test data represents what the operator wants.

For this case study, we want to investigate the (typical) scenario in which the size of the unlabelled dataset is significantly larger than that of the labelled dataset. As such, we constrain the training set and consider here the one vs. all scenario for the ESA Sentinel fleet (Sentinel-1A, Sentinel-1B, Sentinel-2A, Sentinel-2B, Sentinel-3A, Sentinel-3B and Sentinel-5P). From the BAS3E dataset (Section \ref{sec: dataset}), we find 2189 conjunction cases involving members of the Sentinel fleet over a 7 day screening period, which we then combine with an equal number of randomly sampled non-conjunction cases (for which at least one party is a Sentinel) in order to obtain a balanced dataset. This balancing is crucial in ensuring that the trained model does not ignore the outlier conjunction cases from such an intrinsically imbalanced dataset, and is able to successfully capture their defining characteristics. Finally, we include each object pair twice, with its order reversed, in order to prevent biases, and use a nominal 80\% to 20\% random splitting strategy to obtain 7005 object pairs for the training set and 1751 for the validation set.

For the testing set, we randomly sample 10,000 object pairs over the Sentinel operational altitude band (650~km - 850~km), ensuring that no pairs comprise a Sentinel, and again ensuring class balance. Finally, we construct the pre-training dataset in a similar way. For the task of pairwise conjunction screening, the most relevant features are combination (interaction) features over the two objects and thus for this task, we pre-train a pairwise-orbit model (with 12 channels instead of 6) using the same techniques discussed for the single-orbit ORBERT model discussed in Section \ref{sec: orbert}. To aid the generalisation ability of the downstream model, we include the object pairs from both the training and test set in the pre-training set, and randomly sample the remaining pairs over the Sentinel altitude band (ensuring class balance) to obtain a 100,000 object pair pre-training dataset. 


\subsection{Approach}
\label{sec: downstream: approach}

In this work, we phrase the concept of conjunction screening as a machine learning classification task. As such, we aim to train a model to predict a binary class, based on orbital data as an input, as to whether a given object pair will have a conjunction over the next 7 days, such that it can be subsequently filtered or retained for further assessment. We therefore seek to train a time series classifier (with or without the help of the weights of a pretext model) using the training dataset of orbit ephemeris pairs and binary conjunction/no-conjunction labels, and evaluate its performance on unseen test data.

To understand the value of using a pretext model, and the effect of the quantity of pre-training data used, we pre-train two models: the first using only the first 20,000 object pairs, and the second with the full 100,000. For this, we use the same data representation, pre-processing steps, masking parameters and training and architectural hyperparameters as previously discussed in Section \ref{sec: orbert: method} (Tables \ref{tab: mask settings} and \ref{tab: model settings}).

We then train and evaluate the performance of the downstream model (which requires the same architecture as that used by the pre-trained model) under three scenarios: a baseline case, for which the weights of the model are learnt solely from the training data, and two cases where the model is initialised with the weights of the two pre-trained models. 

For this, we again use the hyperparameter settings detailed in Table \ref{tab: model settings}, with the exception of the loss function, which we take as the cross-entropy loss for the new classification task. In each case we allow training of all weights (i.e., no layer of the model is kept frozen) to enable the downstream model to fully adapt to the training data, and train using a flat-cos learning rate schedule as the pre-trained model is already ``warmed'' and thus does not need a warm up. We then train each scenario a total of 10 times.




\subsection{Results}
\label{sec: downstream: results}


\begin{table*}[h!]
  \begin{center}
    \caption{Model performance on validation and testing sets for three training configurations. Performance given as average over 10 training runs using the ROC-AUC metric.}
    \vspace{1em}
    \renewcommand{\arraystretch}{1.2}
    \small 
    \begin{tabularx}{\textwidth}{Xccc} 
      \hline
      & \textbf{Without pre-trained} & \textbf{With pre-trained (20,000)} & \textbf{With pre-trained (100,000)} \\
      \hline
      Valid set & 0.901 +/- 0.006 & 0.901 +/- 0.004 & \textbf{0.905 +/- 0.003} \\
      Test set & 0.684 +/- 0.019 & 0.717 +/- 0.011 & \textbf{0.723 +/- 0.014} \\
      \hline
    \end{tabularx}
    \label{tab:downstream_results}
  \end{center}
\end{table*}


The results of these experiments are given in Table \ref{tab: downstream results}, using the area under the ROC (receiver operating characteristic) curve (AUC) metric. This metric was chosen as it captures the performance of a binary classifier with varying discrimination threshold (rather than assuming a probability threshold of 0.5 is most appropriate), and thus is more descriptive of the ability of the classifier to distinguish between classes than other classical metrics such as accuracy.

From Table \ref{tab: downstream results}, we can draw several conclusions. Firstly, we see that the model performs very well on the validation set (with an AUC score of over 0.9) but, as expected, this performance decreases as we try to generalise to out of distribution data (to the testing set). Most importantly, however, we can see that the performance of this test set not only improves with the addition of pre-trained weights, but that the more unlabelled orbit data we leverage, the better the performance. This effect is present, but much reduced for the validation set, although it should be stressed it is an improvement in the unseen testing data, and not the validation data, that is the aim of this approach. We can therefore conclude that leveraging unlabelled orbit data using a pre-trained model can be a novel and useful methodology in improving the performance of STM tasks, particularly in instances where we generalise to out of distribution data. In this case study, this latter effect can likely be attributed to the fact that the test set is more representative of the pre-training set (as they were constructed from the same distribution), than the biased (Sentinel-centric) training set.


\section{Conclusions \& Future Work}
\label{sec: conclusions}


Inspired by the breakthrough success of BERT-like self-supervised language models in the field of natural language processing, in this work we proposed a new methodology for exploiting large quantities of readily available orbit data to improve the performance of ML-based STM tasks. Our approach is based in the pre-training of an orbit model using self-supervised learning, by masking and reconstructing sections of orbit ephemerides, in order to learn meaningful orbit representations which can be used to aid in a variety of downstream orbit-related tasks.

In this paper, we introduced our pre-trained orbit model, ORBERT, and as a proof of concept of the approach, applied it to the downstream task of all vs. all conjunction screening, phrased here as a time series classification task. Using a cluster analysis of its embedding space, we demonstrated that the orbit model was able to successfully capture physically interpretable representations of different perturbation regimes, without explicitly being trained to do so. This knowledge could then be leveraged for downstream tasks through the use of transfer learning, particularly for cases where the availability of labelled data is limited. We considered the case in which an operator in possession of a limited one vs. all conjunction dataset of their own satellite fleet sought to generalise to predict conjunctions in an all vs. all scenario, and showed that leveraging unlabelled orbit data through this approach lead to a marked improvement in performance, with greater quantities of orbit data providing greater improvements, which hints that the approach scales well to larger datasets.



This new methodology opens a variety of promising avenues for future research. Firstly, we would like to extend the analysis undertaken in this work by investigating the generalisation capability of such pre-trained orbit models to future time periods, as well as any additional gains in performance that can be achieved by further scaling up the quantity of pre-training data. As one of the main advantages of this methodology lies in its flexibility and applicability to a range of different downstream tasks, in future work we will also expand this approach to other STM applications. Possible applications include the use of an orbit-based pre-trained model for manoeuvre detection, to the pre-training of a model using light curve data for aiding in the task of space object classification.

\section*{Acknowledgments}

This research is supported by the EU H2020 MSCA ITN Stardust-R, grant agreement 813644. We gratefully acknowledge the support of NVIDIA Corporation with the donation of a Titan V GPU, used for this research.


\bibliographystyle{acm}
\bibliography{bibliography}



\end{document}